# Transverse momentum as a source of gravitoelectromagnetism

D. H. Delphenich
Spring Valley, OH 45370 USA

**Abstract:** Momentum can be regarded as a "mass current" that can be used as the source of the gravitoelectromagnetic field, which is the weak-field gravitational analogue of the classical electromagnetic field. Typically, gravitoelectromagnetic research considers only mass currents of "convective" type, which are collinear with the velocity vector field, but when one looks at charged fluids that interact with a background electromagnetic field or charged, spinning point-like matter, such as the Frenkel electron, or extended spinning mass distributions, such as the Dirac electron and the Weyssenhoff fluid, one will also encounter transverse momentum. The effects of that transverse momentum on the gravitoelectromagnetic field are investigated.

**1. Introduction.** – One of the most elementary things that one learns in physics is that there is a partial analogy between Coulomb's law of electrostatic attraction and repulsion and Newton's law of universal gravitation. What makes the analogy partial is the fact that to date gravitational repulsion has never been observed, so one always assumes that this also implies that whereas charge can be positive or negative, gravitational mass must always be positive, in order to produce only attraction.

Some early attempts were made to extend Coulomb's law of electrostatics to a more complete law of electrodynamics that would include additional magnetic forces that would originate in the relative motion of the interacting charges. Such an extension was first suggested by Gauss in a note that was published only posthumously in his collected works and then recast in a different form by Wilhelm Weber [**1**] and Franz Neumann [**2**]. As Joseph Bertrand later showed [**3**], the formulas of Gauss and Weber were actually equivalent. One interesting aspect of the Gauss-Weber law was that it included a large parameter that played the role of the speed of propagation of the interaction, which was assumed to be that of light, and as that parameter become infinite, the Gauss-Weber formulas would both go back to that of Coulomb.

These extensions of Coulomb's law eventually led to corresponding extensions of Newton's law, and mostly in the context of celestial mechanics. First, Gustav Holzmüller [**4**] used Weber's law of attraction in a Hamilton-Jacobi approach to planetary motion, and later Tisserand [**5**] used a more direct approach to introducing both Gauss's and Weber's laws. Typically, the effect that researchers in celestial mechanics were looking for was a precession of the longitude of the perihelion of planetary orbits. Oliver Heaviside [**6**] also looked into extending Coulomb's law to a more complete analogy between electromagnetism and gravitation.

Another significant difference between electrostatics and gravitostatics besides the weak analogy between charge and mass is that many orders of magnitude separate the two types of force. That is, the gravitational interaction is so feeble in comparison to the electrostatic interaction that one must typically look for gravitational effects in the context of astrophysics in order to be dealing with the kind of masses that would produce noticeable gravitational fields and the kind of distances at which the effects would assert themselves (e.g., celestial mechanics, stellar interiors, and galactic structure).

In 1918, Hans Thirring published a paper in *die Physikalische Zeitschrift* [**7**] in which he calculated the effect of rotating distant masses in Einstein's theory of gravitation. Basically, he was addressing Mach's suggestion that the collective rotation of the distant stars would also lead to a deformation of the surface of the water in a bucket, just as the rotation of the bucket with respect to the distant stars would. At one point, he included a footnote that proposed that there was a close analogy between the weak-field Einstein equations when the source of the gravitation was a rotating mass and Maxwell's equations of electromagnetism. In a subsequent paper with Josef Lense in *Zeitschrift für Physik* [**8**], Thirring then calculated what it now called the *Lense-Thirring effect* for an elementary orbital scenario. That effect, in its original form, manifested itself as a precession of the plane of an orbit around a gravitating body that would be due to the rotation of the body. In a third paper in Phys. Zeit [**9**], Thirring elaborated upon the formal analogy between the Maxwell's equations and Einstein's equations of gravitation in the linear approximation. There was also a correction to the first paper that showed up in 1921 [**10**] that was based upon some conversations between Thirring and Max von Laue and Wolfgang Pauli.

In essence, the Lense-Thirring effect amounts to the possibility that the rotation of a gravitating body will produce a "dragging of the frames" in its vicinity that will manifest itself as the precession of a gyroscope that is placed in the gravitational field. It is not surprising that many decades passed before the Lense-Thirring effect was actually put to an experimental test. One needs only to recall that the magnetic forces that are produced by an electric current are typically much weaker than the electrostatic ones to see that in the gravitational case, where the analogue of the electrostatic force is already quite weak, the effects of any possible "gravitomagnetic" field that is produced by the motion of a gravitating body, such as its rotation around an axis, would be even more imperceptible. Only the advances of experimental technology made such measurements even possible. It was in 1960 that Leonard Schiff [**11**] first proposed that one could possibly test the Lense-Thirring effect using a satellite in a low Earth orbit. That suggestion, in turn, took a few more decades to implement. The first attempt was with the LAGEOS II satellite program, which was deployed by the Space Shuttle in 1992, and the latest test of frame-dragging was made during the years 2004-2006. The result of the measurement seemed to be positive, although there was considerable debate regarding the actual accuracy of the measurement. During the years 2004-2005, the Gravity Probe B improved upon the accuracy of the LAGEOS II result ([1]); the rate of precession that Gravity Probe B measured was $-37.2 \pm 7.2$ milliseconds of arc per year. It is important to notice that even though the Lense-Thirring effect was first derived from Einstein's theory of gravitation (but in the linear approximation), since the gravitational fields of most celestial bodies, such as the Earth, Sun, and Moon, are actually quite weak in the eyes of general relativity, one would expect that the Maxwellian analogy would be entirely sufficient to describe the phenomenon in most astrophysical problems.

Just as one can seek to extend the scope of Maxwell's equations to something more analogous to Einstein's equations, one can take the opposite approach and look for how

---

([1])  For detailed discussions of the experimental test of the Lense-Thirring effect, as well as more references on the subject, one can confer the survey articles by Bahram Mashoon [**12**] and Ruggiero and Tartaglia [**13**], as well as the book [**14**] by Ignacio Ciufolini and John Archibald Wheeler. Note, however, that the last two references predated the actual satellite experiments.

Einstein's equations might lead to field equations of Maxwell type. The main school of thought in regard to that is based upon the Bianchi identities for the Weyl tensor. Since we will not employ that approach, we shall only refer to some references on the subject [**12**, **15**].

The purpose of this study is to investigate what would be involved with extending the Maxwellian analogy with weak-field gravitation even further by allowing the possibility that some material media are, in a sense, "polarizable" by gravitoelectromagnetic fields in the same way that electric and magnetic dipoles can form in electromagnetic media. In particular, we are mostly concerned with the gravitomagnetic analogue of magnetic dipoles, since the apparent absence of negative masses that would complete the electrostatic analogy would seem to preclude the existence of gravitoelectric dipoles, as well. Our basic thesis is that since energy-momentum can be regarded as a "mass current" that would serve as the source of the gravitoelectromagnetic field, one might consider that in addition to the convective or longitudinal momenta that take the relativistic forms of $m_0$ **u** or $\rho_0$ **u** for point-like and extended masses, respectively, there is such a thing in nature as "transverse" momentum, which can typically originate in the interaction of a charge with a background electromagnetic field or take the form of a relativistic quantum sort of effect that is most definitive in the context of the Dirac electron, which included its spin. One would then naturally wonder about the corresponding gravitomagnetic field that would be produced by the transverse component of the mass current in the same way that one considers polarization currents in the electromagnetic context, as well as the possible physical effects that might be accessible to experimental testing.

The next section of this paper summarizes the relevant notions from Maxwell's theory that will be applied in the gravitoelectromagnetic analogy. The third section then goes over that analogy and discusses the physical interpretation of the basic analogue fields. The fourth section discusses the general concept of transverse momentum and gives various examples of how it can occur naturally in physical models. The fifth section then adds the transverse momentum to the mass current that generates the gravitoelectromagnetic field and examines the consequences. Finally, there is a discussion of the limitations of the present analysis and the possible extensions in its scope.

**2. Maxwellian electromagnetism.** – Before we present the basic analogy between Maxwellian electromagnetism and weak-field gravitation, we shall first summarize the key facts of Maxwellian electromagnetism. We shall first present things in the vector calculus formulation and then in terms of the more modern calculus of exterior differential forms.

*a. Vector calculus formulation of the field equations*. – To many physicists, the most common way of expressing Maxwell's equations for electromagnetism is the form that is found in Jackson [**16**]; in Gaussian units, they are:

$$\nabla \times \mathbf{E} + \frac{1}{c}\frac{\partial \mathbf{B}}{\partial t} = 0, \qquad \nabla \cdot \mathbf{B} = 0, \qquad \nabla \times \mathbf{H} - \frac{1}{c}\frac{\partial \mathbf{D}}{\partial t} = \frac{4\pi}{c}\mathbf{J}, \qquad \nabla \cdot \mathbf{D} = 4\pi\sigma, \qquad (2.1)$$

in which $\mathbf{E}$ is the electric field strength, $\mathbf{D}$ is its corresponding electric excitation (often called the "displacement"), $\mathbf{H}$ is the magnetic field strength, $\mathbf{B}$ is its corresponding excitation (often called the magnetic "flux density"), $\sigma$ is the electric charge density, and $\mathbf{J}$ is the electric current.

When one is dealing with static fields, the time derivatives will drop out, and one will be left with:

$$\nabla \times \mathbf{E} = 0, \qquad \nabla \cdot \mathbf{D} = 4\pi\sigma, \qquad \nabla \times \mathbf{H} = \frac{4\pi}{c}\mathbf{J}, \qquad \nabla \cdot \mathbf{B} = 0, \qquad (2.2)$$

in which the order of equations has been changed in order to show that one now has two sets of decoupled equations for the static electric and magnetic fields.

So far, the system of linear, first-order partial differential equations (2.1) that we have presented for the four spatial vector fields $\mathbf{E}$, $\mathbf{D}$, $\mathbf{H}$, $\mathbf{B}$ is underdetermined, since there are eight equations for twelve component functions when $\sigma$ and $\mathbf{J}$ are given.

One can reduce the twelve components to six independent ones by introducing an *electromagnetic constitutive law* for the medium in which the fields exist, which will take the functional form:

$$\mathbf{D} = \mathbf{D}\,(\mathbf{E},\mathbf{H}), \qquad \mathbf{B} = \mathbf{B}\,(\mathbf{E},\mathbf{H}). \qquad (2.3)$$

Hence, one has six equations that relate twelve functions, which will leave six independent ones. One now has an overdetermined system of eight equations for six unknown functions.

One first notes that a natural consequence of the equations is essentially the conservation of charge:

$$\frac{\partial \sigma}{\partial t} + \nabla \cdot \mathbf{J} = 0. \qquad (2.4)$$

That condition represents an identity that reduces the number of independent equations by one. In order to add one more identity that would make the equations well-determined, one needs to introduce an electromagnetic potential 1-form, which we shall define shortly.

When one is dealing with the classical vacuum, the relationship between the field strengths and the excitations that they produce is the simplest possible one:

$$\mathbf{D} = \varepsilon_0\,\mathbf{E}, \qquad \mathbf{B} = \mu_0\,\mathbf{H}, \qquad (2.5)$$

in which $\varepsilon_0$ is the vacuum dielectric constant, and $\mu_0$ is the vacuum magnetic permeability.

However, in polarizable electromagnetic media, the relationship between field strengths and their corresponding excitations can be much more involved. In particular, $\mathbf{E}$ can produce both electric and magnetic dipoles as a result of its presence in the medium, as can **H**. Indeed, the relationship (2.5) would be typical of media in which no such dipoles are produced. We shall return to elaborate upon this concept at the end of this section.

*b. Field equations in terms of differential forms.* – Since the mid-Nineteenth Century, mostly due to the work of Frobenius, Clebsch, and Darboux on the Pfaff problem, as well as Grassmann's introduction of exterior algebra (but in a different form from its modern definition), the basic ideas of what for a while came to be called the "symbolic calculus" or "absolute calculus" were growing into a replacement for the vector calculus that was more broad-ranging in its applications. Its definition was formalized by Élie Cartan and Edouard Goursat at about the same period of time (viz., 1921). Cartan applied the calculus of exterior differential forms to geometric sorts of things, such as the geometry of moving frames [**17**], while Goursat applied it to more analytical ones, such as the Pfaff problem [**18**], and eventually Cartan, along with Kähler, discussed the integrability of systems of exterior differential equations, which was a generalization of both the Pfaff problem and the Cauchy-Kowalevski theorem for partial differential equations.

Meanwhile, the applications of differential forms to physics were also increasing in number and general familiarity. Such researchers as Godbillion, Souriau, and Gallisot were applying the calculus of differential forms to analytical mechanics (mostly by way of Hamiltonian mechanics), while Debever made one of the early applications of differential forms to general relativity after Cartan. The application of differential forms to Maxwell's theory of electromagnetism was implicit in Minkowski's relativistic formulation of it, but the explicit mention of its relationship to what Cartan and Goursat were doing did not come until later.

This is not the place to introduce the calculus of differential forms to the uninitiated, so we refer them to the books by Henri Cartan [**19**], Walther Thirring [**20**], and Theodore Frenkel [**21**] for that introduction. Hence, we shall start by simply summarizing the formulation of Maxwell's equations in terms of differential forms and then show how the gravitational analogy comes about. For the benefit of the physicists, we shall often give both the "basis-free" form of the geometric objects and their component form in terms of a natural coframe field ($dx^\mu$, $\mu = 0, \ldots, 3$) that is defined by a coordinate system ($U, x^\mu$) on the space-time manifold, which shall be simply Minkowski space $\mathfrak{M}^4 = (\mathbb{R}^4, \eta)$. Our sign convention for the scalar product $\eta$ is that its components in an orthonormal coframe will be:

$$\eta_{\mu\nu} = \text{diag}[+1, -1, -1, -1]. \tag{2.6}$$

Typically, $x^0 = ct$ will be the time coordinate while $\{x^i, i = 1, 2, 3\}$ will be the spatial ones. The line $[x^0]$ in $\mathfrak{M}^4$ that is generated by $x^0$ will be called the *time line*, while the three-dimensional linear space $\Sigma$ that is spanned by the $x^i$ will be called *space*. Space is also the annihilating hyperplane of the 1-form $dx^0$ then. This coordinate system then defines a *time+space splitting* of the space-time – i.e., a direct sum decomposition $\mathfrak{M}^4 = [\mathbf{t}] \oplus \Sigma$ into a one-dimensional time line $[\mathbf{t}]$ and a three-dimensional space $\Sigma$. However, one does not need to define a coordinate system in order to define a time+space splitting ([1]).

One first redefines **E** and **B** as a spatial 1-form and a spatial 2-form, resp.:

$$E = E_i \, dx^i, \qquad B = \tfrac{1}{2} B_{ij} \, dx^i \wedge dx^j. \tag{2.7}$$

The components $B_{ij}$ of the 2-form $B$ relate to the components $B^i$ of the vector field **B** by way of:

$$B_{ij} = \varepsilon_{ijk} B^k, \qquad B^i = \tfrac{1}{2} \varepsilon^{ijk} B_{jk}, \tag{2.8}$$

in which the Levi-Civita symbols $\varepsilon_{ijk}$ and $\varepsilon^{ijk}$ equal $+1$ when $ijk$ is an even permutation of 123, $-1$ when the permutation is odd, and 0 otherwise.

This relationship between the spatial vector **B** and the spatial 2-form $B$ amounts to the spatial version of the *Poincaré isomorphisms* $\#_s : \Lambda_k \Sigma \to \Lambda^{3-k} \Sigma$, which are defined when one chooses a volume element $V_s \in \Lambda^3 \Sigma$ for $\mathbb{R}^3$. Here, we are using $\Lambda_k \Sigma$ to represent the linear space of $k$-vector fields on $\mathfrak{M}^4$ and $\Lambda^k \Sigma$ to represent the linear space of $k$-forms on it. That is, an element **b** of $\Lambda_k \Sigma$ ($k = 0, 1, 2, 3$) will represent a completely-antisymmetric, totally-contravariant tensor field of rank $k$ on $\Sigma$:

$$\mathbf{b} = \frac{1}{k!} b^{i_1 \cdots i_k} \partial_{i_1} \wedge \ldots \wedge \partial_{i_k} \qquad (\partial_i \equiv \frac{\partial}{\partial x^i}), \tag{2.9}$$

while an element $\alpha$ of $\Lambda^k \Sigma$ will represent a completely-antisymmetric, totally-covariant tensor field of rank $k$ on $\Sigma$:

$$\alpha = \frac{1}{k!} \alpha_{i_1 \cdots i_k} \, dx^{i_1} \wedge \ldots \wedge dx^{i_k}. \tag{2.10}$$

The symbol $\wedge$ then represents the *exterior product* in all of this – i.e., the completely-antisymmetrized tensor product.

A volume element $V_s$ on $\Sigma$ is then a non-zero 3-form, such as:

$$V_s = dx^1 \wedge dx^2 \wedge dx^3 = \frac{1}{3!} \varepsilon_{ijk} \, dx^i \wedge dx^j \wedge dx^k. \tag{2.11}$$

For each $k$, the Poincaré isomorphism $\#_s : \Lambda_k \Sigma \to \Lambda^{3-k} \Sigma$ then takes the $k$-vector field **b**, as in (2.9), to the $(3-k)$-form:

$$\#_s \mathbf{b} = i_\mathbf{b} V_s, \tag{2.12}$$

in which $i_\mathbf{b}$ is the *interior product* operator. If **b** is a vector field and the $k$-form $\omega$ takes the form $\alpha_1 \wedge \ldots \wedge \alpha_k$ then:

---

([1]) For more details on the geometry of space-times with time+space splittings (also called 1+3 splittings), one can confer the author's article [**22**] and the references that are cited in it.

$$i_{\mathbf{b}}\omega = \alpha_1(\mathbf{b})\,\alpha_2 \wedge \ldots \wedge \alpha_k - \alpha_2(\mathbf{b})\,\alpha_1 \wedge \alpha_3 \wedge \ldots \wedge \alpha_k + \ldots \pm \alpha_k(\mathbf{b})\,\alpha_1 \wedge \ldots \wedge \alpha_{k-1},$$

or, more concisely:

$$i_{\mathbf{b}}\omega = \sum_{i=1}^{k} (-1)^{i+1} \alpha_i(\mathbf{b})\,\alpha_1 \wedge \ldots \wedge \widehat{\alpha_i} \wedge \ldots \wedge \alpha_k, \qquad (2.13)$$

in which the caret signifies that the 1-form in question has been omitted from the product.

A $k$-form such as $\alpha_1 \wedge \ldots \wedge \alpha_k$ is called *decomposable*. Not all $k$-forms are typically decomposable, and one extends this definition of $i_{\mathbf{b}}$ to all $k$-forms "by linearity." That is, if the $k$-form $\beta$ is a linear combination $\sum_{m=1}^{p} \lambda_m \omega^m$ of $p$ decomposable $k$-forms $\omega^m$ then the interior product $i_{\mathbf{b}}\beta$ will be defined by:

$$i_{\mathbf{b}}\beta = \sum_{m=1}^{p} \lambda_m\, i_{\mathbf{b}}\omega^m. \qquad (2.14)$$

As it happens, since $\Sigma$ is three-dimensional, all $k$-forms on $\Sigma$ will be decomposable, anyway, but the generalizations of these definitions to four dimensions will be useful later on.

In order to extend $i_{\mathbf{b}}$ from vector fields $\mathbf{b}$ to $k$-vector fields, one starts with decomposable $k$-vector fields of the form $\mathbf{b} = \mathbf{b}_1 \wedge \ldots \wedge \mathbf{b}_k$ and an $l$-form $\alpha$ ($l \geq k$) and defines:

$$i_{\mathbf{b}}\,\alpha = i_{\mathbf{b}_1 \wedge \ldots \wedge \mathbf{b}_k}\alpha = i_{\mathbf{b}_k} \cdots i_{\mathbf{b}_1}\alpha. \qquad (2.15)$$

In particular, notice the inversion of the order in the sequence of vectors $\mathbf{b}_1, \ldots, \mathbf{b}_k$.

In the case when $\alpha$ is the non-zero three form $V_s$ on $\Sigma$, if $\mathbf{b}$ represents each of the elements in the sequence $b$, $b^i\,\partial_i$, $\tfrac{1}{2}b^{ij}\,\partial_i \wedge \partial_j$, $b^{123}\,\partial_1 \wedge \partial_2 \wedge \partial_3$, in turn, then $\#_s\mathbf{b}$ will be equal to:

$$b\,V_s, \qquad \tfrac{1}{2}\varepsilon_{ijk}\,b^k\,dx^i \wedge dx^j, \qquad \tfrac{1}{2}\varepsilon_{ijk}\,b^{jk}\,dx^i, \qquad b^{123},$$

respectively.

In particular, from (2.8), we see that the 2-form $B$ relates to the vector field $\mathbf{B}$ by way of:

$$B = \#_s\mathbf{B}. \qquad (2.16)$$

One can assemble the spatial 1-form $E$ and the spatial 2-form $B$ into a 2-form $F$:

$$F = c\,dt \wedge E - B \qquad (2.17)$$

that one calls the *electromagnetic field strength* 2-form. Note that we are now assuming that the components of $E$ and $B$ are also functions of $t$, as well as the spatial variables.

The first set of Maxwell equations in (2.1) can then be absorbed into:

$$d{\wedge}F = 0, \qquad (2.18)$$

in which $d_\wedge : \Lambda^k \mathfrak{M}^4 \to \Lambda^{k+1} \mathfrak{M}^4$ is the *exterior derivative operator* on Minkowski space.

In order to see how that happens, one first substitutes $F$ from (2.17) into (2.18) to first get:

$$d_\wedge F = -c\, dt \wedge d_\wedge E - d_\wedge B.$$

If one temporarily reverts to the local components of $E$ and $B$ then one can verify that:

$$d_\wedge E = dt \wedge \partial_t E + d_{s\wedge} E, \qquad d_\wedge B = dt \wedge \partial_t B + d_{s\wedge} B,$$

in which $d_{s\wedge} : \Lambda^k \Sigma \to \Lambda^{k+1} \Sigma$ is the spatial exterior derivative operator. In particular:

$$d_{s\wedge} E = \tfrac{1}{2}(\partial_i E_j - \partial_j E_i)\, dx^i \wedge dx^j, \qquad (2.19)$$

$$d_{s\wedge} B = \tfrac{1}{3}(\partial_i B_{jk} + \partial_j B_{ki} + \partial_k B_{ij})\, dx^i \wedge dx^j \wedge dx^k. \qquad (2.20)$$

All of this makes:

$$d_\wedge F = -dt \wedge (\partial_t B + c\, d_{s\wedge} E) - d_{s\wedge} B,$$

and since the two terms in this are linearly-independent, the collective vanishing of $d_\wedge F$ would be equivalent to:

$$\partial_t B + c\, d_{s\wedge} E = 0, \qquad d_{s\wedge} B = 0. \qquad (2.21)$$

In order to show that these equations are equivalent to the first two Maxwell equations (2.1), one needs only to show how $\nabla \times \mathbf{E}$ relates to $d_{s\wedge} E$ and how $\nabla \cdot \mathbf{B}$ relates to $d_{s\wedge} B$. In fact, one has:

$$\nabla \times \mathbf{E} = \#_s^{-1} d_{s\wedge} E, \qquad \nabla \cdot \mathbf{B} = \#_s^{-1} d_{s\wedge} B. \qquad (2.22)$$

In order to see that first relationship, it is sufficient to look at the components of $d_{s\wedge} E$, as in (2.19), and then note that:

$$\#_s^{-1}(dx^i \wedge dx^j) = \varepsilon^{ijk} \partial_k. \qquad (2.23)$$

At this point, it helps to point out something that is usually overlooked in purely-mathematical treatments of differential forms, namely, the fact that when an $n$-dimensional manifold $M$ admits a volume element $V$ (hence, it – or rather its tangent bundle – must be orientable), and therefore a set of Poincaré isomorphisms $\# : \Lambda_k M \to \Lambda^{n-k} M$, which are defined as usual by:

$$\#\mathbf{b} = i_\mathbf{b} V, \qquad (2.24)$$

one can define an adjoint operator to $d_\wedge$ by using $\#$, namely:

$$\mathrm{div} = \#^{-1} d_\wedge \#. \qquad (2.25)$$

This operator does, in fact, generalize the divergence of a vector field, since one finds that if $\mathbf{X} = X^\mu \partial_\mu$ is a vector field on $M$ then:

$$\mathrm{div}\, \mathbf{X} = \partial_\mu X^\mu. \qquad (2.26)$$

In particular, one sees that for the spatial vector field **B**, one will have:

$$\text{div}_s \mathbf{B} = \#_s^{-1} d_{s\wedge} \#_s \mathbf{B} = \#_s^{-1} d_{s\wedge} B = \partial_\mu B^\mu,$$

so:

$$d_{s\wedge} B = \#_s \text{div}_s \mathbf{B} = (\partial_\mu B^\mu) V_s. \tag{2.27}$$

One finds that, in general, if **b** is a *k*-vector field of the form (2.9) then the components of div **b** will take the form:

$$(\text{div } \mathbf{b})^{i_2 \cdots i_k} = \partial_{i_1} b^{i_1 i_2 \cdots i_k}. \tag{2.28}$$

Like the operator $d_\wedge$, one can easily show that:

$$\text{div} \cdot \text{div} = 0. \tag{2.29}$$

However, the operator div does not have any distinctive properties in regard to the exterior product, as $d_\wedge$ does; i.e., it is not an anti-derivation.

We are now in a position to deal with the second set of Maxwell equations. One first defines **D** to be a spatial vector field and **H** to be a spatial bivector field and assembles them into the *electromagnetic excitation bivector field* on Minkowski space:

$$\mathfrak{H} = \frac{1}{c} \partial_t \wedge \mathbf{D} + \mathbf{H}. \tag{2.30}$$

If we define the electric charge-current density vector field by way of:

$$\mathbf{j} = \sigma \partial_t + \mathbf{J} \quad (j^t = s, j^i = J^i) \tag{2.31}$$

then we can summarize the second two Maxwell equations – i.e., the source equations – as:

$$\text{div } \mathfrak{H} = 4\pi \mathbf{j}. \tag{2.32}$$

This time, we have defined the space-time volume element to be the non-zero 4-form:

$$V = dx^0 \wedge V_s = dx^0 \wedge dx^1 \wedge dx^2 \wedge dx^3 = \frac{1}{4!} \varepsilon_{\mu_0 \cdots \mu_3} dx^{\mu_0} \wedge \cdots \wedge dx^{\mu_3}. \tag{2.33}$$

Straightforward, but tedious calculations will show that (2.32) is equivalent to:

$$\#_s^{-1} d_{s\wedge} H - \frac{1}{c} \partial_t \mathbf{D} = \frac{4\pi}{c} \mathbf{J}, \qquad \text{div}_s \mathbf{D} = 4\pi \sigma. \tag{2.34}$$

These equations can be compared to the second set of equations in (2.1).

A consequence of (2.32) is derived from (2.29), and takes the form of the conservation of electric charge:

$$\text{div } \mathbf{j} = 0 \qquad (0 = \partial_\mu j^\mu = \partial_t \sigma + \partial_i j^i). \tag{2.35}$$

From the version of the Poincaré lemma that pertains to the div operator, there must exist a bivector field **b** such that:

$$\mathbf{j} = \text{div } \mathbf{b} \tag{2.36}$$

(if only locally).

*c. Potential 1-forms.* – The first Maxwell equation, in the form (2.18), in conjunction with the Poincaré lemma, implies that there is a 1-form:

$$A = c\phi \, dt - A_s, \tag{2.37}$$

such that:

$$F = d_\wedge A. \tag{2.38}$$

One would call such a 1-form an *electromagnetic potential 1-form*. The first Maxwell equation then becomes an identity that follows from the fact that $d_\wedge^2 = 0$ for any possible $A$. However, when one substitutes $d_\wedge A$ for $F$ in the constitutive law, the second Maxwell equation (2.32) will take the form:

$$\text{div } \mathfrak{H} \, (d_\wedge A) = 4\pi \mathbf{j}, \tag{2.39}$$

which represents four equations for the four unknown functions that take the form of the components of $A$. However, since one also has the identity (2.35), the system will become underdetermined by one variable, which can be accounted for by making a choice of a gauge for $A$; i.e., replacing $A$ with $A + d\lambda$, means that the gauge degree of freedom amounts to the free choice of function $\lambda$.

When one does the actual exterior differentiation of $A$ in the form (2.37), one will get:

$$d_\wedge A = c \, d\phi \wedge dt - d_\wedge A_s = - c \, dt \wedge (d\phi + \frac{1}{c}\partial_t A_s) - d_s \wedge A_s.$$

Equating this to $F$ in the form (2.17) will give:

$$E = -(d\phi + \frac{1}{c}\partial_t A_s), \qquad B = d_s \wedge A_s, \tag{2.40}$$

and with the usual identification of the differential operators with their analogues in vector calculus, one will get:

$$\mathbf{E} = -(\nabla \phi + \frac{1}{c}\dot{\mathbf{A}}), \qquad \mathbf{B} = \nabla \times \mathbf{A}, \tag{2.41}$$

which is the way that one would find the association of field strengths and potentials presented in many conventional texts on electromagnetism.

For a topologically-general space-time manifold, a 1-form such as *A* would exist only locally (i.e., on a neighborhood of each point), but for Minkowski space, which is contractible, it will exist globally.  Furthermore, it will not be unique, since one can add any closed 1-form $\chi$ (i.e., $d \wedge \chi = 0$) to *A* and produce another 1-form that will give *F* upon exterior differentiation.  That freedom to alter *A* by a closed 1-form without changing the resulting *F* is referred to as *gauge invariance*, and a choice of *A* or $\chi$ is referred to as a *gauge*.  Typically, the closed form is represented as an exact form (i.e., $\chi = d\lambda$), which is possible locally, in general, and globally in the present case of Minkowski space.

The choice of gauge that will be most interesting to us is the *Lorentz* gauge ([1]), which imposes the condition that the vector field **A** that is metric-dual to the 1-form *A* must have vanishing divergence:

$$\text{div } \mathbf{A} = 0. \tag{2.42}$$

*d. Lorentz force law.* – The force that the combined electric and magnetic fields exerts upon a point-like mass *m* with a charge of *q* that moves with a velocity of **v** relative to the source of the fields is given by the Lorentz law:

$$\mathbf{F} = q\left(\mathbf{E} + \frac{1}{c}\mathbf{v} \times \mathbf{B}\right). \tag{2.43}$$

If one thinks of the motion of *q* as an electric current that takes the form of:

$$\mathbf{J} = q\mathbf{v} \tag{2.44}$$

then the Lorentz force law can also be written:

$$\mathbf{F} = q\mathbf{E} + \frac{1}{c}\mathbf{J} \times \mathbf{B}. \tag{2.45}$$

One can think of the current **J** as it is defined in (2.44) as being of *convective* type; i.e., it is collinear with the velocity, as if *q* were "carried along" by **v**.

The Lorentz force law is even more concisely formulated in terms of the 2-form *F*:

$$f = \frac{1}{c} i_\mathbf{j} F \qquad (f_\nu = \frac{1}{c} j^\mu F_{\mu\nu}). \tag{2.46}$$

In order to see the equivalence of (2.46) with (2.45), substitute (2.17) for *F* and get:

$$i_\mathbf{j}(c\, dt \wedge E) - i_\mathbf{j} B = -(c\, i_\mathbf{j} E)\, dt + c J^t E - i_\mathbf{j} B = -c E(\mathbf{J})\, dt + c q E - i_\mathbf{J} B,$$

---

([1])  Interestingly, this gauge was *not* named after the Dutch physicist Hendrik Antoon Lorentz, but a Russian one by the name of Ludwig Valentin Lorentz, who has been otherwise passed over by the history of gauge field theory.

but since:

$$i_\mathbf{J} B = i_\mathbf{J} \#_s \mathbf{B} = i_\mathbf{J} i_\mathbf{B} V_s = \#_s (\mathbf{B} \wedge \mathbf{J}) = - \#_s (\mathbf{J} \wedge \mathbf{B}),$$

one will get:

$$f = - E(\mathbf{J}) \, dt + q \, E + \frac{1}{c} \#_s (\mathbf{J} \wedge \mathbf{B}). \tag{2.47}$$

The **F** in (2.45) is essentially the spatial part of this, while the temporal component $- E(\mathbf{J})$ represents the power that is being absorbed or radiated by **J**.

*e. The polarization of electromagnetic media.* – Now, let us return to the relationship between the electromagnetic excitation bivector field $\mathfrak{H}$ and the electromagnetic field strength 2-form $F$.

One will have an invertible map $C : \Lambda_2 \mathfrak{M} \to \Lambda^2 \mathfrak{M}$ that is a diffeomorphism of each fiber $\Lambda_{2,x} \mathfrak{M}$ with each fiber $\Lambda^2_x \mathfrak{M}$, and in particular it will take the electromagnetic field strength $F$ to the electromagnetic excitation bivector field:

$$\mathfrak{H} = C(F). \tag{2.48}$$

Such a map is called an *electromagnetic constitutive law*, although in order for $C$ to be an algebraic operator on 2-forms, one must assume that the medium in question is dispersionless ([1]) in both time and space. If there were dispersion, in that sense, then the map $C$ would become an integral operator.

In the component formulation of $C$, we can introduce the "zero-point field" $Z(x) = C(x, 0)$, and characterize the constitutive law $C$ in the component form:

$$[C(F)]^{\mu\nu} = Z^{\mu\nu}(x) + \tfrac{1}{2} C^{\mu\nu\kappa\lambda}(x, F) F_{\kappa\lambda}. \tag{2.49}$$

The maps $C_x : \Lambda_{2,x} \mathfrak{M} \to \Lambda^2_x \mathfrak{M}$ will be linear isomorphisms iff:

$$Z^{\mu\nu}(x) = 0, \qquad C^{\mu\nu\kappa\lambda}(x, F) = C^{\mu\nu\kappa\lambda}(x), \tag{2.50}$$

and that would make $C$ a *linear* electromagnetic constitutive law. The absence of a functional dependency of the components $C^{\mu\nu\kappa\lambda}(x, F)$ on the point $x$ in space-time would make it *homogeneous*, but one should be cautioned that such a property is defined only in specialized frame fields. Indeed, it is only when the transition function $h : \mathfrak{M} \to GL(4)$ is constant that things that are constant in one frame field will be constant in another one.

Although it is possible to discuss Maxwellian electromagnetism without any reference to a space-time metric, nonetheless, if one wishes to compare the constitutive law that is defined by $C$ with the one that is defined by the classical electromagnetic vacuum then one will need to introduce a metric $g$. That will make the association of 2-forms with bivector fields take the form of "raising both indices," namely, the linear isomorphism $\iota_g : \Lambda^2 \mathfrak{M} \to \Lambda_2 \mathfrak{M}$ that takes the 2-form $F = \frac{1}{2} F_{\mu\nu} dx^\mu \wedge dx^\nu$ to the bivector field $\mathbf{F} = \frac{1}{2} F^{\mu\nu} \partial_\mu \wedge \partial_\nu$, such that:

$$F^{\mu\nu} = g^{\mu\kappa} g^{\nu\lambda} F_{\kappa\lambda} = \tfrac{1}{2}(g^{\mu\kappa} g^{\nu\lambda} - g^{\mu\lambda} g^{\nu\kappa}) F_{\kappa\lambda}. \qquad (2.51)$$

---

([1])  Here, we must point out a source of confusion in the theory of electromagnetism, namely, the word "dispersion" is used in two largely-unrelated ways: The present usage refers to the possibility that the state of excitation at a point depends upon its state at neighboring points in space and time. The other usage relates to the way that the frequency of a wave gets coupled to its wave number.

Strictly speaking, this map needs to include $\varepsilon_0$ and $\mu_0$ in order for it to truly represent the electromagnetic constitutive law of the classical vacuum, but we shall ignore that for the moment.

However, it is important to note that since $C$ will define a dispersion law (in the sense of wave motion) that can reduce to a Lorentzian structure in some cases, one must have some way of explaining how the tensor field $g$ (or really the map $\iota_g$) relates to the dispersion law of $C$. Presumably, $\iota_g$ is an asymptotic limit of $C$ that relates to something like the absence of electromagnetic fields.

If that relationship between $\iota_g$ and $C$ is physically meaningful then one can characterize the difference between them by:

$$\mu \equiv C - \iota_g. \qquad (2.52)$$

When this (not-necessarily-invertible) map is applied to the field strength 2-form $F$, the resulting bivector field:

$$\boldsymbol{\mu}(F) = C(F) - \iota_g F \qquad (\mu^{\mu\nu} = Z^{\mu\nu} + \tfrac{1}{2} C^{\mu\nu\kappa\lambda} F_{\kappa\lambda} - F^{\mu\nu}) \qquad (2.53)$$

can be defined to be the *polarization* of the medium that results from its excitation by the field strength $F$.

One can also characterize the difference between $C(F)$ and $\iota_g F$ in terms of the excitation bivector field $\mathfrak{H} = C(F)$:

$$\boldsymbol{\mu}(F) = \mathfrak{H} - \iota_g F \qquad (\mu^{\mu\nu} = \mathfrak{H}^{\mu\nu} - F^{\mu\nu}). \qquad (2.54)$$

Before we go further, it is important to give the concept of the polarization of an electromagnetic medium a more empirical basis. In reality, an electromagnetic medium is polarizable iff the presence of an electromagnetic field $F$ in it provokes the formation of electric or magnetic dipoles (or both). Typically, most conventional media are either dielectric or magnetic of some description (i.e., diamagnetic, paramagnetic, or ferromagnetic), but not both, which is why, for instance, most optical models assume that optical media are dielectrics (if not insulators) that are not magnetically polarizable, and conversely, most magnetic materials are conductors, not dielectrics. We shall return to this picture of polarizability when we attempt to extend gravitoelectromagnetism analogously.

If one has a time+space decomposition of space-time (or really, its tangent bundle), so one can express $\mathfrak{H}$ in the form (2.30) and $\iota_g F$ in the form:

$$i_g F = \frac{1}{c} \partial_t \wedge \mathbf{E} + \mathbf{B}, \tag{2.55}$$

then that will make the polarization bivector field take the time+space form:

$$\boldsymbol{\mu}(F) = \frac{1}{c} \partial_t \wedge \mathbf{P} + \mathbf{M} = \frac{1}{c} \partial_t \wedge (\mathbf{D} - \varepsilon_0 \mathbf{E}) + (\mu_0 \mathbf{H} - \mathbf{B}); \tag{2.56}$$

i.e., the of *electric polarization* vector field **P** and the *magnetization* bivector field **M** are defined by:

$$\mathbf{P} = \mathbf{D} - \varepsilon_0 \mathbf{E}, \qquad \mathbf{M} = \mu_0 \mathbf{H} - \mathbf{B}. \tag{2.57}$$

When one has decomposed $\mathfrak{H}$ into a sum $i_g F + \boldsymbol{\mu}(F)$, one can similarly decompose the divergence of $\mathfrak{H}$:

$$\operatorname{div} \mathfrak{H} = \operatorname{div} i_g F + \operatorname{div} \boldsymbol{\mu}. \tag{2.58}$$

When this is equated to $4\pi \mathbf{j}$, the second of Maxwell's equations can be rewritten in the form:

$$\operatorname{div} i_g F = 4\pi \mathbf{j} - \operatorname{div} \boldsymbol{\mu}, \tag{2.59}$$

which suggests that one can also regard the effect of the polarization of the medium as being something that induces a *polarization current:*

$$4\pi \mathbf{j}_p \equiv -\operatorname{div} \boldsymbol{\mu} \qquad (4\pi j_p^\nu = -\partial_\mu \mu^{\mu\nu}). \tag{2.60}$$

If we take the divergence of both sides of (2.59) then we will see that we still have:

$$\operatorname{div} \mathbf{j} = 0, \tag{2.61}$$

so since:

$$\operatorname{div} \mathbf{j}_p = 0, \tag{2.62}$$

we must always have that not only is the total current:

$$\mathbf{j}_{\text{tot}} \equiv \mathbf{j} + \mathbf{j}_p \tag{2.63}$$

conserved collectively, but its components must be also conserved individually.

With $\boldsymbol{\mu}$ in the form (2.56), the polarization current will decompose into electric and magnetic contributions:

$$4\pi \mathbf{j}_p = -\frac{1}{c} \operatorname{div}(\partial_t \wedge \mathbf{P}) - \operatorname{div} \mathbf{M}.$$

Since **P** and **M** are both spatial tensor fields, one can say that:

$$\frac{1}{c}\,\text{div}\,(\partial_t \wedge \mathbf{P}) = \frac{1}{c}\,\#^{-1}\,d\wedge \#\,(\partial_t \wedge \mathbf{P}) = \frac{1}{c}\,\#^{-1}\,d\wedge i_{\partial_t \wedge \mathbf{P}}\,V = \frac{1}{c}\,\#^{-1}\,d\wedge i_\mathbf{P}\,i_{\partial_t}\,V = \#^{-1}\,d\wedge \#_s \mathbf{P}$$

$$= \#^{-1}\,(dt \wedge \partial_t\,\#_s\mathbf{P} + d\wedge_s\,\#_s\mathbf{P}) = \frac{1}{c}\,\#_s^{-1}\,\partial_t\,\#_s\,\mathbf{P} + \#^{-1}\,d\wedge_s\,\#_s\mathbf{P} = \frac{1}{c}[\dot{\mathbf{P}} - (\text{div}_s\,\mathbf{P})\,\partial_t\,],$$

and

$$\text{div}\,\mathbf{M} = \#^{-1}\,d\wedge \#\mathbf{M} = -\,c\,\#^{-1}\,d\wedge (dt \wedge \#_s \mathbf{M}) = c\,\#^{-1}\,(dt \wedge d\wedge \#_s \mathbf{M}) = c\,\#^{-1}\,(dt \wedge d\wedge_s \#_s \mathbf{M})$$
$$= \#_s^{-1}\,d\wedge_s \#_s \mathbf{M} = \text{div}_s\,\mathbf{M}.$$

That will make:

$$4\pi\,\mathbf{j} = \frac{1}{c}\,[-(\text{div}_s\,\mathbf{P})\,\partial_t + \dot{\mathbf{P}}\,] + \text{div}_s\,\mathbf{M}. \tag{2.64}$$

The temporal component of this is the *polarization charge density*:

$$\sigma_p = -\frac{1}{c}\,\text{div}_s\,\mathbf{P}, \tag{2.65}$$

while the spatial part is the polarization current (properly speaking):

$$\mathbf{j}_s = \frac{1}{c}\,\dot{\mathbf{P}} + \text{div}_s\,\mathbf{M}. \tag{2.66}$$

**3. Gravitoelectromagnetism.** – Theoretically, gravitoelectromagnetism (which we shall abbreviate by the acronym GEM) amounts to a formal analogy between Maxwell's equations for electromagnetism and the weak-field equations for gravitation. In order to justify the association of corresponding fields, we shall first discuss weak-field gravitation in terms of vector calculus, which is often how the static case is discussed in physics, and then recast it in terms of differential forms.

*a. Vector calculus formulation.* – The analogy between Coulomb's law of electrostatic interaction and Newton's law of gravitation is essentially based upon the first pair of static equations in (2.2). One could then make them the equations of a static Newtonian gravitational field by associating both **E** and **D** with the gravitational acceleration field **g** and associating the electric charge density $\sigma$ with the mass density $\rho$. Furthermore, the change in coupling constant amounts to replacing $4\pi$ with Newton's gravitational constant, which we shall denote by $G_0$. That makes:

$$\nabla \times \mathbf{g} = 0, \qquad \nabla \cdot \mathbf{g} = G_0\,\rho. \tag{3.1}$$

The GEM field amounts to the analogue of **H** and **B** that results from the presence of the mass current $\mathbf{j}_s$ in the same way that **B**, and therefore **H**, will result from the presence of the electric current **J**. Hence, we define the GEM field to be $c\boldsymbol{\omega}$, which replaces both **H** and **B**, while $\mathbf{j}_s$ replaces **J** and $G_0 / c$ replaces $4\pi$. Altogether, we have:

$$\mathbf{g} \Leftrightarrow \mathbf{E} = \mathbf{D}, \quad c\,\boldsymbol{\omega} \Leftrightarrow \mathbf{H} = \mathbf{B}, \quad \rho \Leftrightarrow \sigma, \quad \mathbf{j}_s \Leftrightarrow \mathbf{J}, \quad G_0/c \Leftrightarrow 4\pi. \tag{3.2}$$

That gives:

$$\nabla \times \boldsymbol{\omega} = \frac{G_0}{c^2}\mathbf{j}_s, \quad \nabla \cdot \boldsymbol{\omega} = 0 \tag{3.3}$$

in the static case, and if we return to the dynamic case then we will have:

$$\nabla \times \mathbf{g} + \frac{\partial \boldsymbol{\omega}}{\partial t} = 0, \quad \nabla \cdot \boldsymbol{\omega} = 0, \quad \nabla \times \boldsymbol{\omega} - \frac{1}{c^2}\frac{\partial \mathbf{g}}{\partial t} = \frac{G_0}{c^2}\mathbf{j}_s, \quad \nabla \cdot \mathbf{g} = G_0 \rho. \tag{3.4}$$

The fact that the coupling constant between the source current $\mathbf{j}_s$ and its field $\boldsymbol{\omega}$ is $G_0/c^2$ shows how feeble the strength of that field will be and why the existence of $\boldsymbol{\omega}$ was observed experimentally only recently; in MKS units, $G_0/c^2 = 0.74 \times 10^{-27}$ m/kg.

*b. GEM in terms of differential forms.* – In order to put (3.4) into the language of differential forms, we simply alter the basic analogy (3.2) to take the form:

$$F \Leftrightarrow G = c\,(dt \wedge g - \omega), \quad \mathfrak{H} \Leftrightarrow \mathfrak{G} = -\frac{1}{c}\partial_t \wedge g - c\,\omega, \quad j = \rho\,\partial_t + \mathbf{j}_s. \tag{3.5}$$

In effect, the constitutive law that couples $G$ to $\mathfrak{G}$ is defined by the Lorentzian metric isomorphism $\iota_\eta$, although one should note that with the sign convention that we chose for $\eta$ in (2.6), the spatial 1-form $g$ will go to what we are calling $-\mathbf{g}$, while the spatial 2-form $\omega$ will go to $\boldsymbol{\omega}$.

The field equations will then be:

$$d \wedge G = 0, \quad \text{div}\,\mathfrak{G} = G_0\,j, \quad \mathfrak{G} = i_\eta G, \tag{3.6}$$

or in component form:

$$\partial_\lambda G_{\mu\nu} + \partial_\mu G_{\nu\lambda} + \partial_\nu G_{\lambda\mu} = 0, \quad \partial_\mu \mathfrak{G}^{\mu\nu} = G_0\,j^\nu, \quad \mathfrak{G}^{\mu\nu} = \tfrac{1}{2}(\eta^{\mu\kappa}\eta^{\nu\lambda} - \eta^{\mu\lambda}\eta^{\nu\kappa})G_{\kappa\lambda}. \tag{3.7}$$

Substituting the time+space forms of $G$ and $\mathfrak{G}$ will give:

$$\partial_t\,\omega + d_s \wedge g = 0, \quad d_s \wedge \omega = 0, \quad \text{div}_s\,\mathbf{g} = G_0\,\rho, \quad -\partial_t\,\mathbf{g} + c^2\,\text{div}_s\,\boldsymbol{\omega} = G_0\,\mathbf{j}_s, \tag{3.8}$$

or in component form:

$$\partial_t\,\omega_j + \tfrac{1}{2}(\partial_i\,g_j - \partial_j\,g_i) = 0, \quad \partial_i\,\omega_{jk} + \partial_j\,\omega_{ki} + \partial_k\,\omega_{ij} = 0, \tag{3.9}$$

$$\partial_i\,g^i = G_0\,\rho, \quad -\partial_t\,g^{ij} + c^2\,\partial_i\,\omega^{ij} = G_0\,j^{\,j}, \tag{3.10}$$

which should be compared with (3.4).

In the static case, (3.8) and (3.9), (3.10) will take the form:

$$d_s \wedge g = 0, \qquad \text{div}_s\, \mathbf{g} = G_0\, \rho, \qquad d_s \wedge \omega = 0, \qquad \text{div}_s\, \boldsymbol{\omega} = \frac{G_0}{c^2} \mathbf{j}_s \qquad (3.11)$$

and

$$\partial_i\, g_j - \partial_j\, g_i = 0, \qquad\qquad \partial_i\, g^i = G_0\, \rho, \qquad (3.12)$$

$$\partial_i\, \omega_{jk} + \partial_j\, \omega_{ki} + \partial_k\, \omega_{ij} = 0, \qquad c^2\, \partial_i\, \omega^{ij} = G_0\, j^{\,j}, \qquad (3.13)$$

resp., which can be compared to (3.1) and (3.3).

The first equation in (3.6) implies the existence of a (local) potential 1-form:

$$u = c\, \psi\, dt - u_s\,. \qquad (3.14)$$

Like its electromagnetic analogue, it will not be unique, but will be subject to the same gauge freedom.

Taking the exterior derivative of $u$ will give:

$$d \wedge u = c\, d\psi \wedge dt - d \wedge u_s = -c\, dt \wedge (\partial_t\, u_s + d_s \psi) - d_s \wedge u_s\,.$$

Equating corresponding terms in the right-hand side of $G$ in (3.5) will yield:

$$g = -\partial_t\, u_s - d_s\, \psi, \qquad\qquad \omega = \frac{1}{c} d_s \wedge u_s\,, \qquad (3.15)$$

and in terms of components that will take the form:

$$g_i = -\partial_t\, u_i - \partial_i\, \psi, \qquad \omega_{ij} = \frac{1}{c}(\partial_i\, u_j - \partial_j\, u_i). \qquad (3.16)$$

Hence, since $g$ is a linear acceleration and $\omega$ is an angular velocity, we would be dimensionally correct in identifying $\psi$ with an acceleration potential, while $u_s$ is the spatial part of a proper-time covelocity 1-form. Since the fields $g$ and $\omega$ are also defined at the points where there is no mass to be accelerating or rotating, one must clarify what the values of those fields would represent. Basically, in the absence of gravitomagnetism, $g$ would be the free-fall acceleration of any non-zero mass that is placed at the point. Presumably, the angular velocity $\omega$ would then represent a sort of "free-rotation" speed for a "gravitomagnetic" dipole that is placed at the point. (We shall clarify that statement shortly.)

*c. Force on a moving mass.* – The gravitational analogue of the Lorentz force law comes about by making the replacements (3.2) in (2.43), but with $m$ instead of $\rho$ and $m\,\mathbf{v}$ instead of $\mathbf{j}_s$, which gives:

$$\mathbf{F} = m\,(\mathbf{g} + \mathbf{v} \times \boldsymbol{\omega}), \qquad (3.17)$$

which can be expressed in the relativistic form:

$$f = i_{\mathbf{j}}\, G \qquad (f_\nu = j^\mu\, G_{\mu\nu}). \tag{3.18}$$

One then sees that the contribution to the force that comes from the gravitomagnetic field $\omega$ is essentially that of a Coriolis force (except for the missing factor of 2).

Note that even though the forces that are imparted by the electrostatic field **E** of an individual charge often overshadow those of its magnetostatic field **B**, nonetheless, one must notice that industrial electromagnets are more commonly used for heavy lifting and magnetic levitation than industrial electrets. In other words, the collective effect of many elementary magnetic dipoles seems to be more useful in practice than that of many individual electric charges. Hence, reasoning by analogy, one should not dismiss the possible empirical significance of the gravitomagnetic field out of hand.

We shall return to a discussion of the meaning of "gravitomagnetic" dipoles later, but first we need to examine the GEM analogue of polarization current, namely, transverse momentum, and give some examples of how it shows up in theoretical physics.

**4. Transverse momentum.** – The most common way of defining momentum in classical mechanics is the convective form of that dynamical quantity. In the case of a non-relativistic point-mass $m$ that moves along a curve in space with a velocity vector **v**, one defines the non-relativistic momentum vector by:

$$\mathbf{P} = m\, \mathbf{v}. \tag{4.1}$$

One then sees that the momentum vector will be collinear with the velocity vector, by definition. This can also be referred to as "longitudinal" momentum.

If the mass is extended over a finite spatial volume and described by a mass density $\rho$, while the motion is defined by a congruence of spatial curves with a velocity vector field **v** then one can define a corresponding momentum density vector field:

$$\mathbf{p} = \rho\, \mathbf{v}. \tag{4.2}$$

Once again, this would be a convective or longitudinal definition of momentum density.

Similarly, when one goes on to relativistic mechanics, the main alterations that are necessary are that the curve or congruence of curves in space that are parameterized by the time coordinate $t$ must become a time-like world-line or congruence of time-like world-lines in space-time, resp., that are parameterized by proper time $\tau$, so the spatial velocity vector field **u** must become a time-like vector field **u** on space-time, in which typically the derivative is with respect to proper time. When a world-line has been parameterized by proper-time, the effect of that is to make:

$$u^2 = \eta(\mathbf{u},\mathbf{u}) = c^2. \tag{4.3}$$

Finally, the mass *m* or mass density $\rho$ must become a proper mass $m_0$ or proper mass density $\rho_0$, resp., and one defines the energy-momentum four-vector for a point-like mass and the energy-momentum density four-vector for an extended mass by:

$$\mathbf{P} = m_0 \mathbf{u} \quad \text{and} \quad \mathbf{p} = \rho_0 \mathbf{u}, \tag{4.4}$$

respectively.

Actually, since one can think of dynamics as being dual to kinematics by way of the virtual work functional, it is often more appropriate to think of **P** or **p** as a spatial or time-like space-time 1-form *P* or *p*, resp., that relates to the vector fields by way of the spatial or space-time metric, resp.; i.e., one lowers the index on its components. That puts the last two equations into the forms:

$$P = m_0 u \quad \text{and} \quad p = \rho_0 u, \tag{4.5}$$

resp., in which the 1-form *u* is the metric dual of **u** (i.e., $u_\mu = \eta_{\mu\nu} u^\nu$) and is referred to as *covelocity*.

However, this convective or longitudinal scenario is not the only one possible. Typically, in situations in which the matter in question is charged and interacting with a background electromagnetic field or just spinning, there will be contributions to the momentum (or energy-momentum) that are not collinear with velocity, and those contributions are then referred to as *transverse momentum*. By definition, if the energy-momentum density vector field ([1]) takes the form:

$$\mathbf{p} = \rho_0 \mathbf{u} + \mathbf{p}_t \tag{4.6}$$

then $\mathbf{p}_t$ will represent transverse momentum iff:

$$\eta(\mathbf{u}, \mathbf{p}_t) = 0. \tag{4.7}$$

If one wishes to express this in terms of the corresponding energy-momentum density 1-form:

$$p = \rho_0 u + p_t \tag{4.8}$$

then one will have:

$$i_\mathbf{u} p_t = p_t(\mathbf{u}) = 0. \tag{4.9}$$

Note that as a result:

$$p(\mathbf{u}) = \rho_0 u(\mathbf{u}) = \rho_0 c^2, \tag{4.10}$$

so the transverse momentum makes no contribution to the rest energy density if is defined in this way. However:

$$p^2 = \eta(\mathbf{p}, \mathbf{p}) = (\rho_0 c)^2 + (\mathbf{p}_t)^2 = (m_{\text{eff}} c)^2, \tag{4.11}$$

with:

---

([1]) Although we are making this definition for the relativistic case, the non-relativistic definition is entirely analogous.

$$m_{\text{eff}} = [(\rho_0)^2 + (\mathbf{p}_t / c)^2]^{1/2} = \rho_0 \left[1 - \left(\frac{p_t}{\rho_0 c}\right)^2\right]^{1/2}. \tag{4.12}$$

(The minus sign appears due to the fact that $\mathbf{p}_t$ is a space-like vector.)

One might distinguish the notion of transverse momentum from an earlier one that Lorentz exhibited in his theory of the electron [**23**] that takes the form of longitudinal and transverse mass. Basically, the Fitzgerald-Lorentz contraction of a mass distribution that is spherical in its rest space into an (apparent) oblate spheroid due to its relative motion will imply that, in effect, the mass will be different in the longitudinal direction from what it is in the directions that are in the plane perpendicular to velocity. In particular, if the rest mass is $m_0$ and the relative speed is $\beta = v / c$ then the longitudinal and transverse masses will be:

$$m_l = \frac{m_0}{(1 - \beta^2)^{3/2}}, \qquad m_t = \frac{m_0}{(1 - \beta^2)^{1/2}}, \tag{4.13}$$

respectively.

One might imagine a mass matrix ([1]) $M_{ij} = \text{diag}\,[m_l, m_t, m_t]$ that takes spatial velocities to spatial momenta, but one has to realize that such a matrix is defined in a frame that is adapted to the velocity vector, so the transverse component of velocity would be zero, by definition, as would the corresponding transverse momentum.

The examples of transverse momentum that we shall examine are all relativistic ones that typically relate to relativistic quantum wave mechanics, namely, the minimal electromagnetic coupling of an external electromagnetic field to energy-momentum, the Frenkel electron, the Dirac electron, and the Weyssenhoff fluid.

*a. Minimal electromagnetic coupling.* – When a point-like charged mass (rest mass = $m_0$, charge = $q$) moves in the presence of an external electromagnetic field $F = d{\wedge}A$, one can also absorb the external electromagnetic field into the definition of the charge-field system by way of *minimal electromagnetic coupling:*

$$P = m_0 u - \frac{q}{c} A. \tag{4.14}$$

One will then have:

$$P(\mathbf{u}) = m_0 c^2 - \frac{q}{c} A(\mathbf{u}), \tag{4.15}$$

and as for $P^2 = (m_{\text{eff}}\, c)^2$, one will have:

$$m_{\text{eff}} = m_0 \left[1 - \frac{2q}{m_0 c^3} A(\mathbf{u}) + \left(\frac{q}{m_0 c^2}\right)^2 A^2\right]. \tag{4.16}$$

---

([1]) Mass matrices are used in the context of the electroweak model for particle interactions, but in a different sense than the present one.

If one expresses $A$ in time+space form as in (2.37) and expresses $\mathbf{u}$ as $\gamma[(1/c)\,\partial_t + \mathbf{v}]$, with $\gamma = (1 - v^2/c^2)^{-1/2}$, then:

$$A(\mathbf{u}) = \gamma[\phi - A_s(\mathbf{v})]. \tag{4.17}$$

Note that in a rest frame ($\mathbf{v} = 0$, $\gamma = 1$), the only contribution to $A(\mathbf{u})$ will come from the electrostatic potential. Since $A(\mathbf{u})$ does not have to vanish unless $\phi = A_s(\mathbf{v})$, $A$ does not have to be purely transverse, although typically it will contain both longitudinal and transverse components.

One can see that the Lorentz force law becomes equivalent to the conservation of total energy-momentum:

$$0 = \frac{dP}{d\tau} = m_0\,\dot{u} - \frac{q}{c}\frac{dA}{d\tau}. \tag{4.18}$$

That is because:

$$\frac{dA}{d\tau} = L_{\mathbf{u}}A = i_{\mathbf{u}}d{\wedge}A + d{\wedge}(i_{\mathbf{u}}A) = i_{\mathbf{u}}F + d\,[A(\mathbf{u})] = i_{\mathbf{u}}F. \tag{4.19}$$

The term $d\,[A(\mathbf{u})]$ vanishes because $\mathbf{u}(\tau)$ is defined only as a function of $\tau$, so the same thing will be true of the scalar $A(\mathbf{u})$, and therefore its differential must vanish.

Hence, (4.18) will take the form:

$$m_0\,\dot{u} = \frac{q}{c}\,i_{\mathbf{u}}F. \tag{4.20}$$

Things are quite different for a charged, extended mass, such as a charged fluid. If its proper mass density is $\rho_0$, its charge density $\sigma$, and its proper-time-parameterized flow velocity is $\mathbf{u}$ then $\mathbf{u}(x)$ will be function of $\tau$ only implicitly by way of its functional dependency upon the space-time point $x$ when one selects a trajectory $x(\tau)$. The energy-momentum density 1-form $p$ will then take the form:

$$p = \rho_0\,u - \frac{\sigma}{c}A. \tag{4.21}$$

Now, when one takes the proper-time derivative of $p$, the result will contain terms that depend upon the differentials of $\rho_0$ and $\sigma$, as well:

$$\frac{dp}{d\tau} = \frac{d}{d\tau}(\rho_0 u) - \frac{\dot{\sigma}}{c}A - \frac{\sigma}{c}\frac{dA}{d\tau}. \tag{4.22}$$

Now:

$$\dot{\sigma} = L_{\mathbf{u}}\sigma = \mathbf{u}\,\sigma = u^\mu\,\partial_\mu\,\sigma, \tag{4.23}$$

and

$$\frac{dA}{d\tau} = L_{\mathbf{u}}A = i_{\mathbf{u}}F + d\,(A(\mathbf{u})), \tag{4.24}$$

so the vanishing of $dp/d\tau$ would imply that:

$$\frac{d}{d\tau}(\rho_0 \, u) = \frac{\sigma}{c} i_\mathbf{u} F + \frac{1}{c}[(\mathbf{u} \, \sigma) \, A + \sigma d \, (A(\mathbf{u}))]. \tag{4.25}$$

Note that even for proper mass and charge densities that are constant in time and space, there will still be an extra contribution to the analogue of (4.20) that comes from $d \, (A(\mathbf{u}))$; i.e.:

$$\rho_0 \dot{u} = \frac{\sigma}{c} i_\mathbf{u} F + \frac{\sigma}{c} d \, (A(\mathbf{u})). \tag{4.26}$$

*b. Frenkel electron.* – In 1929, one year after Dirac presented his relativistic theory of the spinning electron, Joseph Frenkel made a first attempt [**24**] at a relativistic theory of a charged, spinning electron that interacted with an external electromagnetic field $F$. His model still assumed a point-like distribution of mass, charge, and spin for the electron, but allowed the point to rotate. Admittedly, this sounds more like a simplifying approximation, since rotation is more naturally defined in terms of extended matter, but eventually this way of thinking came to be regarded as the "pole-dipole" approximation to an extended distribution, which can be thought of as a Lorentzian (i.e., orthonormal) frame moving along a time-like world-line.

Without going into all of the details, we shall simply summarize the relevant consequences of Frenkel's model. The time-like velocity vector field of the world-line is **u**, while its corresponding covelocity 1-form is $u$, the proper mass is $m_0$, and the spin is described by a 2-form $S$ that satisfies the "Frenkel constraint":

$$i_\mathbf{u} S = 0 \qquad (u^\mu S_{\mu\nu} = 0). \tag{4.27}$$

This has the effect of making the motion of the Lorentzian frame purely rotational with respect to a rest frame and is based in the fact that experiments suggested that although the electron has an intrinsic magnetic dipole moment in its rest space, nonetheless, it seems to have no electric dipole moment. The electromagnetic dipole moment 2-form $\mu$ for the spinning electron is coupled to that spin 2-form by way of the "Uhlenbeck-Goudsmit hypothesis":

$$\mu = -\mu_B \, S, \tag{4.28}$$

in which $\mu_B = e\hbar/2m_e c$ (in CGS units) is the Bohr magneton.

The energy-momentum 1-form $p$ that one derives for this motion is:

$$p = m_\text{eff} \, u - \frac{1}{c^2} i_\mathbf{a} S, \tag{4.29}$$

into which we have introduced an *effective mass:*

$$m_\text{eff} = m_0 + \frac{1}{2c^2} F(\boldsymbol{\mu}), \tag{4.30}$$

which includes a contribution from the potential energy of the coupling of the electromagnetic dipole moment to the external electromagnetic field and a vector field **a** along the world-line whose corresponding metric-dual 1-form is:

$$a = \frac{g}{2}\frac{e}{m_0 c} i_{\mathbf{u}} F - \dot{u}. \tag{4.31}$$

The $g$ in this is a constant that characterizes the state of rotation. In particular, for non-spinning charges, $g = 2$.

Although **a** clearly has the dimensions of an acceleration, it represents a measure of the difference between the force law that applies to a spinning charge and the usual Lorentz force law. In particular, for non-spinning charges the Lorentz force law would apply, which would make $\mathbf{a} = 0$. Otherwise, one would have a variation on the Lorentz force law that would take the form:

$$\dot{p} = \frac{e}{c} i_{\mathbf{u}} F + \tfrac{1}{2} dF(\mathbf{\mu}). \tag{4.32}$$

Hence, there is an additional force that acts upon the spinning charge by way of the coupling of its electromagnetic dipole moment **μ** to the inhomogeneity $dF$ in the external electromagnetic field. (Recall that in the Stern-Gerlach experiment in order to exhibit the spin of the electron, it was necessary to employ an inhomogeneous magnetic field.)

If one equates $(e / c)\, i_{\mathbf{u}} F$ in this with the corresponding expression for it that one infers from (4.31) and solves for **a** then one will get:

$$a = \frac{g}{2m_0}[\dot{p} - \tfrac{1}{2} dF(\mathbf{\mu})] - \dot{u}. \tag{4.33}$$

Note that even in the absence of an external field [$F = 0$, so $m_{\text{eff}} = m_0$, $a = (g/2m_0)\dot{p} - \dot{u}$], one will still have a contribution from **a** when $g$ is not equal to 2; i.e., when the charged mass is spinning.

In any event, the transverse momentum will take the form:

$$p_t = -\frac{1}{c^2} i_{\mathbf{a}} S. \tag{4.34}$$

*c. Dirac electron*. – When Paul Dirac published his quantum theory of the electron in 1928 [**25**], one of the big obstacles that it faced was the esoteric and largely unfamiliar nature of its introduction of the Clifford algebra of Minkowski space $\mathcal{C}(4, \eta)$, at least as far as the rest of the physicists of the era were concerned. As a result of that introduction, the way that classical observables were "encrypted" into the Dirac wave function was not entirely unique or agreed upon. Typically, what the various attempts to derive classical (but relativistic) observables from the Dirac wave function $\Psi$ had in common was that they generally started with the definition of the sixteen "bilinear covariants."

One defines the bilinear covariants of $\Psi$ by first defining a representation of the sixteen-real-dimensional Clifford algebra $\mathcal{C}(4, \eta)$ in the sixteen-complex-dimensional algebra of 4×4 complex matrices $M(4, \mathbb{C})$ (although the choice of that representation is not at all uniquely agreed-upon by physics). Four generators of $\mathcal{C}(4, \eta)$, namely, a choice of Lorentzian frame $\{\mathbf{e}_\mu, \mu = 0, \ldots, 3\}$, go to four generators of its image in $M(4, \mathbb{C})$, which will be four matrices $\{\gamma_\mu, \mu = 0, \ldots 3\}$. A basis $\{E_A, A = 1, \ldots, 16\}$ for $\mathcal{C}(4, \eta)$ can be defined by all linearly-independent products of the $\mathbf{e}_\mu$, and they will then correspond to a basis $\{\gamma_A, A = 1, \ldots, 16\}$ for the image of $\mathcal{C}(4, \eta)$ in $M(4, \mathbb{C})$, which will be a proper subspace of $M(4, \mathbb{C})$ as a real vector space.

Each basis vector $E_A$ then associates $\Psi$ with a real number $\overline{\Psi} E_A \Psi$ that is the $A^{th}$ *bilinear covariant* of $\Psi$. In this expression, we are defining the "Dirac conjugate" wave function $\overline{\Psi}$ by way of:

$$\overline{\Psi} = \Psi^\dagger \gamma_0, \tag{4.35}$$

in which $\Psi^\dagger$ is the Hermitian conjugate of $\Psi$ (i.e., the complex conjugate of its transpose as a column matrix).

Since $\Psi$ takes its values in $\mathbb{C}^4$, which has a real dimension of eight, and there are sixteen bilinear covariants, one sees that the latter cannot all be algebraically independent. In fact, the algebra of $\mathcal{C}(4, \eta)$ imposes nine identities upon the bilinear covariants, which were first observed by Louis de Broglie and expanded upon more rigorously by Wolfgang Pauli and his student Koffink. Actually, that only leaves seven independent covariants, and the way that one finds the eighth one is to go on to the "differential covariants." We shall not go into the details of that here, but refer the curious to the paper of Takahiko Takabayasi [**26**] in which he discusses the conversion of the Dirac equation into a set of relativistic equations of motion for a relativistic spinning fluid that corresponds to the Dirac wave function. The author's own thoughts on the topic are discussed at length in his book on continuum-mechanical models for wave mechanics [**27**].

The essential fact that we shall cite here is that Takabayasi's expression for the energy-momentum four-vector amounts to:

$$\mathbf{p} = \rho_0 \cos \theta \, \mathbf{u} + \mathrm{div}\, \boldsymbol{\sigma} + i_{\mathrm{grad}\, \theta} *\boldsymbol{\sigma}. \tag{4.36}$$

The bivector field $\boldsymbol{\sigma}$ is the spin tensor that one obtains from $\Psi$, while the somewhat mysterious angle $\theta$ amounts to a sort of phase angle in a plane that is spanned by $\mathbf{u}$ and the vector field $\mathbf{s} = \#^{-1}(u \wedge \sigma)$, which is essentially the Pauli-Lubanski vector field.

The first term in the right-hand side of (4.36) is clearly the convective part of the momentum, while the last term is reminiscent of the transverse momentum in the Frenkel electron. It is the second term – namely, div $\boldsymbol{\sigma}$ – that will be most interesting to us in the next section, because it relates to a sort of "polarization current" that gets associated with the convective current by the formation of spin dipoles in the same way that the polarization of an electromagnetic medium by an electromagnetic field is associated with a polarization current that takes the form of the divergence of a bivector field.

This decomposition of energy-momentum into a convective part and a polarization current is analogous to the "Gordon decomposition" [28] of the conserved Noether current that is associated with the Dirac electron due to the invariance of its action functional under gauge transformations.

*d. Weyssenhoff fluid.* – The so-called *Weyssenhoff fluid* [29] is essentially a simplification of the Dirac electron, in the sense that the energy-momentum-stress tensor for the Weyssenhoff fluid includes only the kinetic term of the Dirac electron, but not the internal stresses. The Weyssenhoff fluid is defined by a rest mass density $\rho_0$, a time-like velocity vector field **u** for a congruence of world-lines (usually referred to as a "world-tube"), and a spin 2-form $\sigma$ that satisfies the Frenkel constraint (4.27).

As it happens, the form that the energy-momentum density 1-form $p$ takes the form:

$$p = \rho_0 \, u - \frac{1}{c^2} i_\mathbf{a} \sigma, \qquad (4.37)$$

in which **a** is then the proper acceleration – i.e., $d\mathbf{u} / d\tau$. This is essentially (4.29), with $m_{\text{eff}}$ replaced with $\rho_0$, and the spin 2-form $S$ replaced with the spin density 2-form $\sigma$.

Clearly, $\rho_0 \, u$ is longitudinal, while the Frenkel constraint insures that:

$$p_t = -\frac{1}{c^2} i_\mathbf{a} \sigma \qquad (4.38)$$

is, in fact, transverse; i.e.:

$$\eta \, (\mathbf{p}_t, \mathbf{u}) = i_\mathbf{u} \, p_t = -\frac{1}{c^2} i_\mathbf{u} i_\mathbf{a} \sigma = \frac{1}{c^2} i_\mathbf{a} i_\mathbf{u} \sigma = 0.$$

The aforementioned energy-momentum stress tensor takes the simple form:

$$T = p \otimes \mathbf{u}, \qquad (4.39)$$

and if one substitutes (4.37) then that will take the form:

$$T = \rho_0 \, u \otimes \mathbf{u} - \frac{1}{c^2} i_\mathbf{a} \sigma \otimes \mathbf{u}. \qquad (4.40)$$

The first term on the right-hand side is the kinetic term that one expects for a relativistic fluid without spin, so the second term represents a contribution to the internal stresses that is solely due to the presence of spin.

The doubly-covariant form of $T$ is:

$$T = \rho_0\, u \otimes u - \frac{1}{c^2} i_{\mathbf{a}} \sigma \otimes u, \tag{4.41}$$

and when this is polarized into a symmetric and an antisymmetric part:

$$T = T_+ + T_-, \tag{4.42}$$

one will get:

$$T_+ = \rho_0\, u \odot u - \frac{1}{c^2} i_{\mathbf{a}} \sigma \odot u, \qquad T_- = -\frac{1}{c^2} i_{\mathbf{a}} \sigma \wedge u = p_t \wedge u. \tag{4.43}$$

This shows that the asymmetry in the tensor $T$ is solely due to the presence of transverse momentum. Furthermore, since:

$$i_{\mathbf{a}} u = u(\mathbf{a}) = \eta(\mathbf{u},\mathbf{a}) = \tfrac{1}{2}\frac{d}{d\tau} u^2 = 0,$$

one will have:

$$T_- = -\frac{1}{c^2} i_{\mathbf{a}} (\sigma \wedge u) = -\frac{1}{c} i_{\mathbf{a}} \#\mathbf{S} = \frac{1}{c}\#(\mathbf{a} \wedge \mathbf{S}). \tag{4.44}$$

*e. The conservation of mass.* – If one thinks of momentum as basically a mass current then the divergence of that current will say whether the current has a source or not. The relativistic form of the vanishing of the divergence is:

$$0 = \operatorname{div} \mathbf{p} = \partial_\mu p^\mu = \frac{1}{c^2} \partial_t\, \varepsilon + \partial_i\, p^{\,i}. \tag{4.45}$$

In the event that the relativistic energy density $\varepsilon$ takes the form $\rho c^2$, in which $\rho$ represents the relative mass density, the latter law can be expressed in the form:

$$\partial_t\, \rho = -\partial_i\, p^{\,i}, \tag{4.46}$$

which is typical of the balance of mass.

If the vector field **u** that is associated with the motion of $\rho$ is regarded as a flow velocity vector field then one can regard div **p** as the *dynamical incompressibility* of the flow, as opposed to div **u**, which is its *kinematical incompressibility*.

By the dual of the Poincaré lemma, if **p** is dynamically incompressible then there will exist a bivector field **s** such that:

$$\mathbf{p} = \operatorname{div} \mathbf{s}. \tag{4.47}$$

Note that if **p** has units of momentum then **s** must have units of angular momentum.

Of course, **s** is not defined uniquely, since adding any bivector field with vanishing divergence to **s** will not affect **p**. Hence, that is essentially the dual of gauge invariance.

**5. Combining transverse momentum with GEM.** – If we go back to the basic analogy (3.2) then we will see that in effect we were assuming that the analogy between Maxwellian electromagnetism and weak-field gravitation does not extend to the possibility that some regions of the space-time manifold might "polarize" in the presence of a GEM field. That then translates into the idea that the map that takes the 2-form *G* to the bivector field $\mathfrak{G}$ is simply the metric isomorphism $\iota_g : \Lambda_2 \to \Lambda^2$ that amounts to raising both indices of the components.

However, if we wish to go back over the definitions in section 2.*e* and give them gravitational analogues then in order to give that analogy any physical reality, we would really have to start by discussing what the gravitational analogues of electric and magnetic dipoles would be. That is where one must recall that the original analogy between Coulomb's law and Newton's law was not complete to begin with. In particular, to date, no one has truly resolved the issue of what a negative gravitational mass would have to represent, since the interaction of unlike gravitational masses would presumably produce a force of repulsion, just as the interaction of like masses are known to attract.

One finds that naively if a negative gravitational mass also implied a negative inertial mass *à la Galileo* then the effect of the mutual repulsion of a positive and negative mass would be to make both of them accelerate in the *same* direction and chase each other off to infinity in that direction, which sounds highly unphysical. In order to make things seem more realistic, one could assume that the equivalence of gravitational and inertial mass is only true in absolute value, while inertial masses are always positive. Nevertheless, in the absence of direct experiments, one can only speculate in the name of theory.

Relativistic quantum mechanics has to deal with the possibility of negative mass in the form of antimatter, since the Dirac wave functions that correspond to antiparticles have negative kinetic energy eigenvalues. Hence, there is probably good reason to at least speculate on the possibility that equal and opposite masses that are separated by a finite distance might represent mass dipoles that would be analogous to electric ones. Indeed, the virtual pairs of particles and their antiparticles would then seem to play a dual role in the name of vacuum polarization in both the electric and gravitational contexts.

The concept of a gravitational analogue of a magnetic dipole is probably easier to justify physically, since the gravitational analogue of magnetism has much in common with spin, in its quantum sense. Hence, one might suspect that the effect of a strong enough gravitomagnetic field on a material medium that possesses spin degrees of freedom might be to bring about the alignment (i.e., polarization) of those spin dipoles. The fact that neutrons have non-zero spin tends to suggest that perhaps the interiors of neutron stars might be a likely place to find such effects.

*a. General construction for incorporating transverse momentum.* – So far, we have really introduced two basic sources of transverse momentum that typically have complementary domains of definition: Spin degrees of freedom, which only exist at the points of space-time where the source of a gravitational field exists, and the electromagnetic potential 1-form *A*, which is defined wherever an electromagnetic field is defined. However, in the latter case, the effect of multiplying *A* by the electric charge density $\sigma$ in order to get an energy-momentum means that the spatial support of $\sigma A$ will be contained in the spatial support of $\sigma$. Hence, we shall examine each of these in turn as possible contributions to the source current of the GEM field equations with understanding the changes will pertain to the points at which the source current is defined.

First, let us assume that the mass current **j** that serves as the source of the GEM field $\mathfrak{G}$ in (3.6) is composed of a convective part and a transverse part:

$$\mathbf{j} = \rho_0 \, \mathbf{u} - \text{div } \mathbf{s}, \tag{5.1}$$

in which bivector field **s** might represent possible spin degrees of freedom in the source mass-current.

In order for div **s** to constitute a transverse momentum, we must have:

$$0 = i_u \, (\text{div } \mathbf{s}) \qquad (0 = u_\nu \, \partial_\mu \, s^{\mu\nu}). \tag{5.2}$$

Since:

$$\partial_\mu \, (u_\nu \, s^{\mu\nu}) = \tfrac{1}{2} (\partial_\mu \, u_\nu - \partial_\nu \, u_\mu) \, s^{\mu\nu} + u_\nu \, \partial_\mu \, s^{\mu\nu},$$

the condition for div **s** to be transverse will also take the form:

$$\text{div } (i_u \, \mathbf{s}) = (d\wedge u)(\mathbf{s}) \quad [\partial_\mu \, (u_\nu \, s^{\mu\nu}) = \tfrac{1}{2} (\partial_\mu \, u_\nu - \partial_\nu \, u_\mu) \, s^{\mu\nu}], \tag{5.3}$$

and if **s** satisfies the Frenkel constraint then the condition for transversality will reduce to:

$$(d\wedge u)(\mathbf{s}) = 0 \qquad [(\partial_\mu \, u_\nu - \partial_\nu \, u_\mu) \, s^{\mu\nu} = 0]. \tag{5.4}$$

When the general expression (5.1) for **p** is substituted in (3.6), the resulting form for that equation will then be:

$$\text{div } i_\eta G = G_0 \, (\rho_0 \, \mathbf{u} - \text{div } \mathbf{s}), \tag{5.5}$$

or

$$\text{div } \mathfrak{G} = G_0 \, \rho_0 \, \mathbf{u} \qquad (\partial_\mu \, \mathfrak{G}^{\mu\nu} = G_0 \, \rho_0 \, u^\nu), \tag{5.6}$$

in which we have replaced the constitutive law in (3.6) with:

$$\mathfrak{G} = \iota_\eta G + G_0 \, \mathbf{s} \qquad (\mathfrak{G}^{\mu\nu} = G^{\mu\nu} + G_0 \, s^{\mu\nu}). \tag{5.7}$$

Hence, the spin-dependent part of the source current can be absorbed into the constitutive law for the material medium by assuming that it is spin-polarizable.

To summarize, the equations of gravitoelectromagnetism are now:

$$d\wedge G = 0, \qquad \text{div } \mathfrak{G} = G_0 \, \rho_0 \, \mathbf{u}, \qquad \text{div } (\rho_0 \, \mathbf{u}) = 0, \qquad \mathfrak{G} = \iota_\eta G + G_0 \, \mathbf{s} \tag{5.8}$$

or

$$\partial_\lambda \, G_{\mu\nu} + \partial_\mu \, G_{\nu\lambda} + \partial_\nu \, G_{\lambda\mu} = 0, \qquad \partial_\mu \, \mathfrak{G}^{\mu\nu} = G_0 \, \rho_0 \, u^\nu, \tag{5.9}$$

$$\partial_\mu (\rho_0 u^\mu) = 0, \qquad \mathfrak{G}^{\mu\nu} = G^{\mu\nu} + G_0\, s^{\mu\nu}, \tag{5.10}$$

in component form.

Now, let us consider the possibility that the source current **j** has both mass density $\rho_0$ and charge density $\sigma$, such as charged fluid, and that it is in the presence of a background electromagnetic field $F$ that is described by a 1-form $A$, whose corresponding metric-dual vector field is **A**. Hence, the electromagnetically-coupled energy-momentum vector field will be:

$$\mathbf{p} = \rho_0\, \mathbf{u} - \frac{\sigma}{c}\mathbf{A}. \tag{5.11}$$

In order to put (5.11) into the same form as (5.1), we can take advantage of the fact that **A** is not defined uniquely, but we would now need to introduce a modification of the Lorentz gauge that includes $\sigma$, this time:

$$0 = \mathrm{div}\,(\sigma \mathbf{A}) = \mathbf{A}\sigma + \sigma\,\mathrm{div}\,\mathbf{A},$$

or

$$\mathrm{div}\,\mathbf{A} = -\frac{1}{\sigma}\mathbf{A}\sigma \qquad (\partial_\mu A^\mu = -\frac{1}{\sigma} A^\mu \partial_\mu \sigma). \tag{5.12}$$

Hence, from the Poincaré lemma for div, there will be a bivector field **b** such that:

$$\sigma \mathbf{A} = \mathrm{div}\,\mathbf{b}. \tag{5.13}$$

**b** will not be unique, either, since one can add any bivector field with vanishing divergence to it and not change the resulting vector field $\sigma \mathbf{A}$.

We now have:

$$\mathbf{p} = \rho_0\, \mathbf{u} - \frac{1}{c}\mathrm{div}\,\mathbf{b}. \tag{5.14}$$

Hence, this time, $(1/c)\,\mathbf{b}$ plays essentially the same role that **s** did before.

We can now repeat what we did in (5.7) and define:

$$\mathfrak{G} = \iota_\eta G + \frac{G_0}{c}\mathbf{b} \qquad (\mathfrak{G}^{\mu\nu} = G^{\mu\nu} + \frac{G_0}{c} b^{\mu\nu}). \tag{5.15}$$

Therefore, with the choice of gauge (5.12) for **A**, we can put the minimally-coupled energy-momentum 1-form into the form (5.1) by simply defining:

$$\mathbf{s} = \frac{1}{c}\mathbf{b}. \tag{5.16}$$

The main difference between the two types of transverse momentum is then simply a matter of physical interpretation. In the case of matter with spin, the transverse contribution to the energy-momentum vector field **p** comes from the divergence of the spin, while in the minimally-coupled electromagnetic case of a charged fluid, the contribution from $(\sigma/c)\,\mathbf{A}$ represents something that has more to with the internal stresses that are developed in the fluid by the interaction of the charge distribution with the background electromagnetic field. Of course, one should recall that momentum flux and stress (i.e., pressure) have the same basic physical units, along with energy density.

*b. Space+time form of the equations.* – In order to put the basic equations of GEM – namely, (5.8) – into space-time form, we first recall that:

$$G = c\,(dt \wedge g - \omega), \qquad \iota_\eta G = -\frac{1}{c}\partial_t \wedge \mathbf{g} - c\,\boldsymbol{\omega}.$$

If we define the bivector field $\mathbf{s}$ in terms of the spatial vector $\boldsymbol{\alpha}$ and the spatial bivector field $\boldsymbol{\varpi}$ as:

$$G_0\,\mathbf{s} = \frac{1}{c}\partial_t \wedge \boldsymbol{\alpha} + c\,\boldsymbol{\varpi} \qquad (5.17)$$

then the constitutive law (5.7) will take the form:

$$\mathfrak{G} = -\frac{1}{c}\partial_t \wedge (\mathbf{g} - \boldsymbol{\alpha}) - c\,(\boldsymbol{\omega} - \boldsymbol{\varpi}). \qquad (5.18)$$

Hence, we are basically replacing $\mathbf{g}$ with $\mathbf{g} - \boldsymbol{\alpha}$ and $\boldsymbol{\omega}$ with $\boldsymbol{\omega} - \boldsymbol{\varpi}$. If those substitutions are made in the third and fourth of equations (3.8) then that will give:

$$\operatorname{div}_s \mathbf{g} = G_0\,\rho + \operatorname{div}_s \boldsymbol{\alpha}, \qquad -\partial_t \mathbf{g} + c^2 \operatorname{div}_s \boldsymbol{\omega} = G_0\,\mathbf{j}_s + \partial_t \boldsymbol{\alpha} + c^2 \operatorname{div}_s \boldsymbol{\varpi}. \qquad (5.19)$$

The first two of equations (3.8) do not change because they relate to the 2-form $G$, not the bivector field $\mathfrak{G}$.

If the spin bivector field $\mathbf{s}$ is purely rotational then $\boldsymbol{\alpha} = 0$, and the last set of equations will take the form:

$$\operatorname{div}_s \mathbf{g} = G_0\,\rho, \qquad -\partial_t \mathbf{g} + c^2 \operatorname{div}_s \boldsymbol{\omega} = G_0\,\mathbf{j}_s + c^2 \operatorname{div}_s \boldsymbol{\varpi}. \qquad (5.20)$$

Hence, the equation for $\operatorname{div}_s \mathbf{g}$ has not changed, but the equation for $\boldsymbol{\omega}$ has picked up another source current from then non-vanishing of $\operatorname{div}_s \boldsymbol{\varpi}$.

*c. Static fields.* – In the static case, the time derivatives will vanish, and equations (5.19) will take the form:

$$d_{s\wedge} g = 0, \qquad \operatorname{div}_s \mathbf{g} = G_0\,\rho + \operatorname{div}_s \boldsymbol{\alpha}, \qquad (5.21)$$

$$d_{s\wedge} \omega = 0, \qquad \operatorname{div}_s \boldsymbol{\omega} = \frac{G_0\,\rho_0}{c^2}\mathbf{u}_s + \operatorname{div}_s \boldsymbol{\varpi}, \qquad (5.22)$$

which can be compared with (3.11). Clearly, the only difference is in the addition of a source mass density in the form of $(1/G_0)$ div$_s$ $\alpha$ and a source mass current in the form of $(c^2 / G_0 \rho_0)$ div$_s$ $\varpi$.

In the purely rotational case, they will further reduce to:

$$d_{s\wedge}g = 0, \qquad \text{div}_s \, \mathbf{g} = G_0 \, \rho, \qquad d_{s\wedge}\omega = 0, \quad \text{div}_s \, \boldsymbol{\omega} = \frac{G_0 \, \rho_0}{c^2} \mathbf{u}_s + \text{div}_s \, \varpi, \quad (5.23)$$

in which the equations for **g** have returned to the Newtonian ones, while the equations for $\omega$ now contain a contribution to the source current that is due to the non-vanishing of div$_s$ $\varpi$.

**6. Discussion.** – With the proliferation of experimentally-untested, if not untestable, theories in physics nowadays, one should probably be even more conscientious about looking for possible ways to test one's theory experimentally. In the case of gravitoelectromagnetic effects, one must typically be dealing with strong gravitational fields or exceptionally-high-precision measuring devices. Since the topic of this study has been the possibility of introducing transverse momentum into the source of a gravitational field – such as spin dipoles – one obvious place to look for such effects would be in the fields of neutron stars, which have both strong gravitational fields and spin dipoles, if they indeed deserve to use the word "neutron" in their name. Hence, a next step to take in the theory would be to look for possible effects that the spin density of neutrons might imply that might serve as the basis for astrophysical observations.

A possible extension in the scope of the present analysis is based in the fact that even in 1918, Thirring [**10**] had already observed that if Maxwell's equations of electromagnetism are analogous to the weak-field gravitational equations then that might suggest that there are some strong-field equations of electromagnetism that would be analogous to Einstein's equations of gravitation. Indeed, he also suggested that such equations would become relevant in the realm of strong electromagnetic field strengths, such as one would find in the close proximity to elementary charges and magnetic dipoles.

Consequently, one would expect that the growing acceptance of GEM as an empirical fact might imply corresponding changes to Maxwellian electromagnetism that would come from the strong-field theory of gravitation. One might even consider that the original form that Einstein gave to his equations might not be the best one for exhibiting the electromagnetic analogy, but simply the best that one could do in the early Twentieth Century, when the dominant approach to the geometry of curved spaces was Riemannian geometry. In subsequent decades, many directions in non-Riemannian geometry have been explored, especially as far as the introduction of non-zero torsion is concerned. Hence, finding the strong-field equations of electromagnetism that would be analogous to strong-field equations of gravitation in the same way that Maxwell's equations are analogous to weak-field gravitation might involve first finding a formulation of the strong-field gravitational equations that would make the analogy more natural.

____________